\documentclass[
journal=jcpc, 
manuscript=article]{achemso}

\usepackage[version=3]{mhchem} 
\usepackage{epsfig}
\usepackage{amsmath}
\usepackage{amssymb}
\usepackage{color}
\usepackage{setspace}
\usepackage{comment}
\usepackage{subfig}
\usepackage{indentfirst}
\usepackage{caption}
\usepackage{float} 
\usepackage{wrapfig} 
\usepackage{graphicx}
\usepackage{subfig}
\usepackage{tikz}
\usepackage{verbatim}
\usetikzlibrary{decorations.pathmorphing}
\usetikzlibrary{arrows}

\usepackage{caption}
\newcommand{\ket}[1]{$\left|#1\right\rangle$}

\newcommand{\onlinecite}[1]{\hspace{-1 ex} \nocite{#1}\citenum{#1}} 
\author{Ravindra Nanguneri}
\author{John Parkhill}
\affiliation{Department of Chemistry and Biochemistry, The University of Notre Dame du Lac \\251 Nieuwland Science Hall Notre Dame, IN 46556}
\email{jparkhil@nd.edu}

\title[\texttt{achemso} demonstration]
{Relaxation Between Bright Optical Wannier Excitons in Perovskite Solar Absorber CH$_3$NH$_3$PbI$_3$}

\begin{document}

\begin{abstract}
We study the light-absorbing states of the mixed-halide perovskite CH$_{3}$NH$_{3}$PbI$_2$Cl and tri-iodide perovskite CH$_{3}$NH$_{3}$PbI$_3$ with density functional and many-body calculations to explain the desirable photovolatic features of these materials. The short-lived electron-hole bound states produced in this photovoltaic material are of halide to lead electron transfer character, with a Wannier-type exciton. Bethe-Salpeter (GW+BSE) calculations of the absorption cross section reveal strong screening of the electron-hole interaction. The atomic character of the exciton retains ligand-to-metal character within the visible spectrum, with differing degrees of localization outside the unit cell. The average electron-hole separation in the lowest exciton is found to be about 5$A^{\circ}$, slightly larger than the Pb-I bond length. Finally, we determine the role of methylammonium's dipole in the ultrafast relaxation by preparing an atomistic model of the picosecond electronic dynamics in the tri-iodide, PbI$_3$. Our model allows us to identify phonon modes which couple strongly to the electronic excitations, and explain the picosecond timescale intra-band relaxation dynamics seen in recent transient absorption experiments. We largely substantiate the conjectured three band model for the dynamics, but also identify other possible relaxation channels in the tri-iodide.\\
%
%
\begin{figure}
\centering
\includegraphics[width=0.25\textwidth]{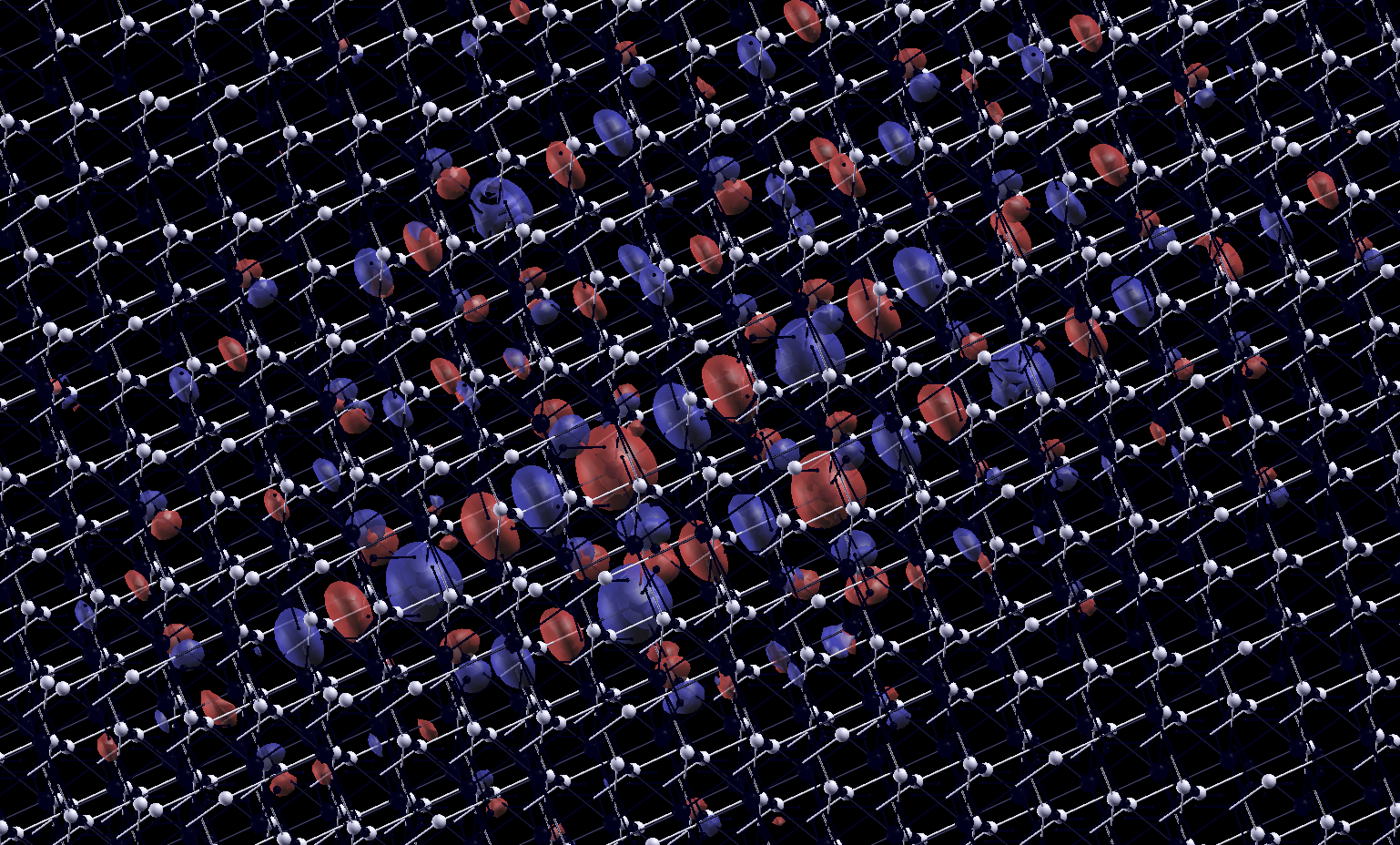}
\label{fig:abstract_fig}
\end{figure}
\end{abstract}

\textit{Introduction.} The organolead methylammonium halide family of perovskite materials can be used to achieve solar light conversion efficiencies of around 15\% \cite{snaith_perovskites_2013,snaith_EESci_2013}, and anode-cathode voltages of about 1.0 volt \cite{liu_efficient_2013}. These materials are prepared much more easily than most solid state cells via a low-temperature solution process, and are readily integrated into layered photovoltaic devices with a variety of junction materials\cite{Malinkiewicz,Kelly}. The significant processing advantages of these materials over silicon are paired with surprisingly robust absorption and charge transport features that have caught the interest of the solar-cell research community\cite{Bretschneider:2014aa,Edri:2013aa,Baikie:2013aa,doi:10.1021/ja411014k,leadfree,Green}. Among materials in this class, the tri-iodide with the formula CH$_{3}$NH$_{3}$PbI$_{3}$ has one of the highest efficiencies, and it is this particular material that we study in this work. \\
\indent The mobility of carriers in the tri-iodide material is on the order of 10$\text{cm}^2\text{V}^{-1}\text{s}^{-1}$, remarkably large for a solution processed material \cite{mob_snaith}. This high mobility is attributed to slower recombination for carriers in the tri-iodide material, and a very small effective mass for electrons in the conduction band\cite{Xing18102013,C4TA01198H}. Dielectric screening of the electron-hole interaction leads to small electron-hole binding energies in this material on the order of 32-100meV\cite{Savenije:0aa,Zhang:2013aa}, and spontaneous thermal dissociation of the exciton at room temperature\cite{DaInnocenzo:2014aa}. The tri-iodide pervoskite has a diffusion length of 100nm\cite{Xing18102013}, favorable for efficient charge transport. The bandstructures of this material also reveal nearly similar electron and hole dispersions near the frontier bands which result in ambipolar transport characterisitics. This leads to nearly equal electron and hole diffusion lengths, another favorable characteristic which in turn implies that the photocurrent is not suppressed by the space charge buildup.\\
%
\indent     Many of the key transport features related to the band structures of these materials have been reproduced with calculations. Kohn-Sham GGA calculations yield qualitatively correct band dispersion and wavefunctions, but underestimate the optical gap. The non-local screened exchange added in GW calculations opens the gap, which is then closed to some extent following treatment of spin-orbit coupling\cite{PhysRevB.89.155204,Umari:2014aa,marcus}. However, the ultrafast dynamics of these materials exhibit several features which are difficult to rationalize on the basis of the band structure alone. \\
\indent    In this paper we study the optical absorption spectrum of the tri-iodide perovskite, using DFT and the Bethe-Salpeter equation,  and we characterize the excited states that are optically active. We also study the role of dynamics of the organic group in the relaxation of the excited states at room temperature. As a preliminary to our central results on the spectrum and excited states' dynamics, we first present DFT+PBE results and verify the known structure, band-structure, and DOS. We find that the absorption spectrum matches well with experiment. By examining the electron and hole density plots of the exciton states, we are able to see the low-lying exciton states have predominant halide p to lead p charge-transfer character and that the average electron-hole separation in these states is around 5\AA, slightly larger than the halide-lead bond length. \\
\indent    We are most interested in the femto- and picosecond relaxation processess between these excited states, and their role in preserving the energy harvested by these materials. It is known experimentally and predicted theoretically that excitons spontaneously dissociate in perovskites at room temperature, but the timescale of this process is not well-established. Photoluminescence of these materials reflects a broad separation of rates for interband and intraband relaxation. Electron- hole recombination is extraordinarily slow (100s of nanoseconds)\cite{C4CC04973J}, but relaxation to the smallest gap states occurs on a picosecond timescale. To unravel the chemical motifs responsible for these valuable dynamical features, we produce estimates of the picosecond dynamics based on a second-order system-bath model of electronic relaxation in perovskites. Our analysis relates the observed transient absorption (TA) spectrum to the free-carrier states. We largely support the assignment of the 400fs kinetics to relaxation from excitons made of hole states below the gap, and the persistence of bound excitons on the picosecond timescale. \\
\indent Our study has focused on both the Cl-doped mixed-halide, as well as the pure tri-iodide perovskites. However, recent consensus among experimentalists is that the Cl doping in the presumed mixed-halide perovskite is not actually present\cite{MGratzel2014}. In this paper, we present the absorption spectrum for the Cl-doped mixed-halide, and the relaxation rates for both the mixed-halide and pure tri-iodide perovskites. \\
%
\textit{Mean-Field.}\indent Periodic KS-DFT calculations were performed using the Quantum Espresso\cite{QE-2009} package. We used the PBE functional, with a 4x4x4 k-grid, a 50/250 Ry wave-function/charge-density cutoff, and the scalar-relativistic MT set of pseudo-potentials to describe the valence-core interactions\cite{MTpseudo}. We performed a structural relaxation of the primitive cell on the initial structure, which was a tetragonal unit cell with a=b=11.2 Bohr, and a distortion $\frac{c}{a}=0.96$. The methylammonium moiety sits near the origin, with the C-N bond axis oriented along the (1,1,1) direction. The organic group is not covalently bound to the inorganic matrix. Steric hindrance prevents rapid inverting rotations, but rotation around the CN-axis occurs readily. In our model, lead occupies the body center of the unit cell, while the halides occupy the face-center positions. Upon optimization, the Pb-I bond lengths are about 3.1 \AA, similar to the theoretical values in Ref.\onlinecite{Mosconi_first-principles-no-soc_2013}. The structure of the optimized primitive cell with basis atoms is shown in Fig. 1(b). \\
\indent The band structure and DOS (Fig. SF1 and SF2)  predicted by our work agrees with previous reports Ref.\onlinecite{Mosconi_first-principles-no-soc_2013}. We found a direct band gap of about 1.55 eV at the R point. The PDOS reveals that the hole states are predominantly composed of I 5p orbitals, while the electron states are composed of the Pb 6p orbitals. Thus, the low-lying KS excitation creates an electron-hole pair where an electron hops from one of the halides to the central lead atom in an octahedron unit. As mentioned in the introduction, incorporation of SOC reduces the band-gap by about 1.0 eV\cite{Even_SOC_calc_2013}, while GW corrections are expected to increase the gap by 1.0 eV\cite{Umari:2014aa}, so that the DFT+SOC+GW gap would be almost identical to the gap neglecting both those effects, and in agreement with the experimental gap. Our main interest is in the relaxation dynamics of the exciton in the material, which would be too expensive to obtain using band energies at the GW, or even SOC, levels of theory. The good reproduction of the relative shape of the spectrum supports the idea that the Kohn-Sham bands behave qualitatively like bands which include SOC and screening effects which would be too costly to treat with dynamics.\\

\begin{figure}
\centering
\includegraphics[width=0.7\textwidth]{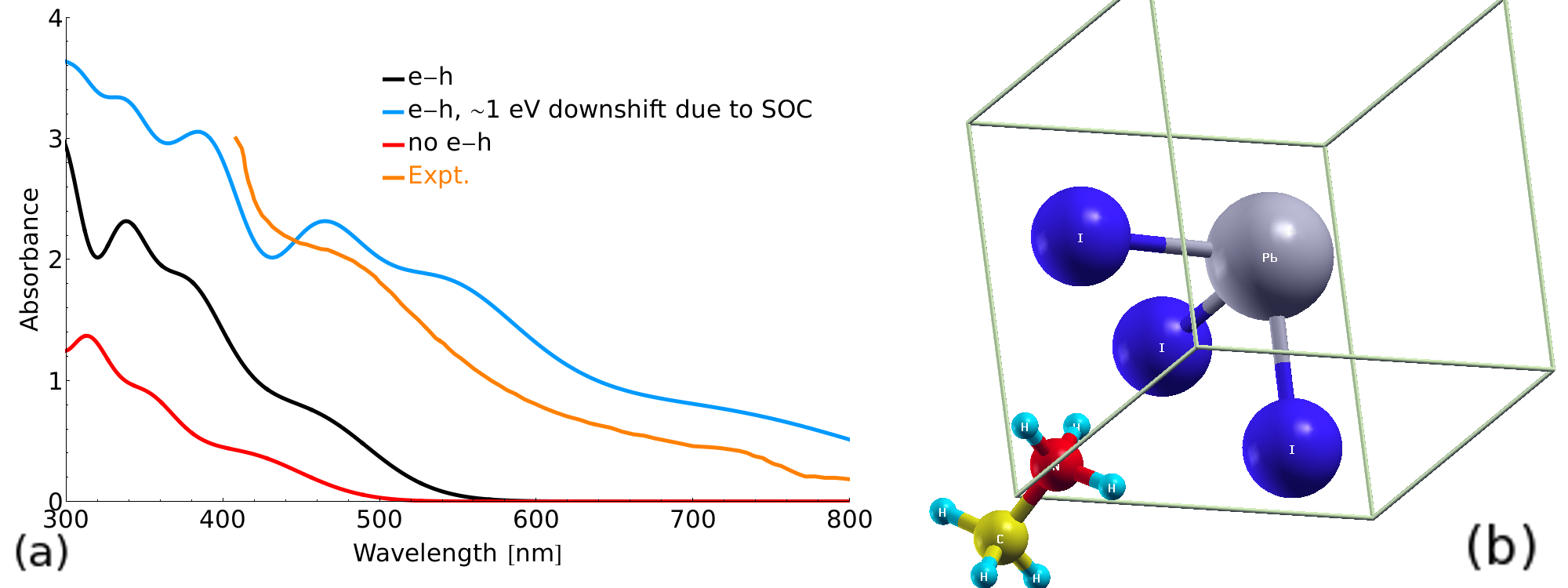}
\caption{(a) Absorption spectra with and without electron-hole interaction for the mixed-halide perovskite CH$_3$NH$_3$PbI$_2$Cl. The BSE interacting spectrum without SOC is calculated with 24 conduction and 12 valence states in the active space. We also show the experimental spectra from Ref.\cite{Lee02112012}, and the BSE spectra shifted to account for SOC. (b) The primitive cell and crystal structure of the tri-iodide perovskite. }
\label{fig:fig1}
\end{figure}

\indent The BSE method is one of the most rigorous affordable theories for the excited states of solids, and includes correlation physics via the screening of the exchange interaction within the so-called GW approximation to the electron self-energy\cite{PhysRev.139.A796,PhysRevB.34.5390,PhysRevB.62.4927}.
We use the single-shot G0W0 and the generalized plasmon-pole approximation as implemented in BerkeleyGW\cite{Deslippe20121269}. Within this approximation, the G0W0 corrected direct gap is found to be about 2.66 eV, also at the R point (1/2,1/2,1/2). The screening and electron-hole interaction effects incorporated in GW+BSE do not qualitatively alter the picture of the optically excited states implied by the Kohn-Sham bandstructure. The electron radial density distribution of the lowest excitonic state, Fig. 2(a), shows the density peaks at a position slightly larger than the lead-halide bond-length, supporting the above view of the electron charge-transfer character. From this plot, we infer that the length scale of the exciton is roughly 5\AA. Unlike the KS bands, the excitons are not eigenstates of the momentum operator, they are localized in real-space, as seen in the electron and hole density distributions in Figure 2(b) and 2(c). The oscillations in the phase between primitive cells are the main features which distinguish the different excitons distributed in the visible spectrum of this material. There is a predominant Pb 6p-orbital character to the electron density, while there is a Iodine 5p-orbital and some Pb 6s-orbital character to the hole density, indicating, in agreement with the aperiodic cluster calculations (Figures SF3(a) and SF3(b) of the Supplement), that the excited electron state mostly resides on the Pb atoms, while the hole state is on the halides and Pb. \\

\indent The BSE method can be viewed as an effective Schrodinger equation for electron-hole pairs, allowing the exciton binding energy to be associated with the difference between the energies of optical states calculated with and without electron-hole interaction (Figure 1). The exciton binding energy $E_{b}$, defined as the difference $E_b=E_{GW}-E_{BSE}$, is about 200 meV. That is less than ambient thermal energy, and within an order of magnitude of experimental estimates of 50-100meV \cite{dinnocenzo_excitons_2014}. Our calculations assume a dielectric function for screening determined by the ground state of the material, whereas the pumped perovskite's dielectric function is likely much larger. It is known that the population of carrier states increases the screening experienced by excitons (the screening due to excited states), decreasing their binding energy\cite{PhysRevB.4.2567}. This effect, which is not included in the BSE, makes perfect agreement of our calculated binding energy with the experimental value difficult. In Figure 1(a), we plot the GW+BSE absorption along with the experimental, non-interacting, and the GW+BSE absorption shifted by ~1 eV that we estimate to be the effect of the neglected SOC.
We find that after accounting for such a shift, the BSE absorption matches quite well with the experimental result for CH$_{3}$NH$_{3}$PbI$_{3}$.
\indent Although the GW-BSE approach is perhaps the best computable method to study the excited states of solids, it is still incapable of representing excited states which are dominated by promotion of multiple electrons into the conduction band. For that we would need correlated wave-function methods such as CISD or CCSD.

\indent We tried to quantify the corrections introduced by inclusion of correlations using RI-SOS-ADC(2) cluster calculations\cite{QChem2014}. We find that the excited state energies are brought down by about -0.02au (=-0.5eV) from their CIS values, and that the weight of the double excitations is very small for the lowest three excited states; the squared norm of the doubles amplitude for these states is 0.06. \\
\begin{figure}
\centering
\includegraphics[width=0.5\textwidth]{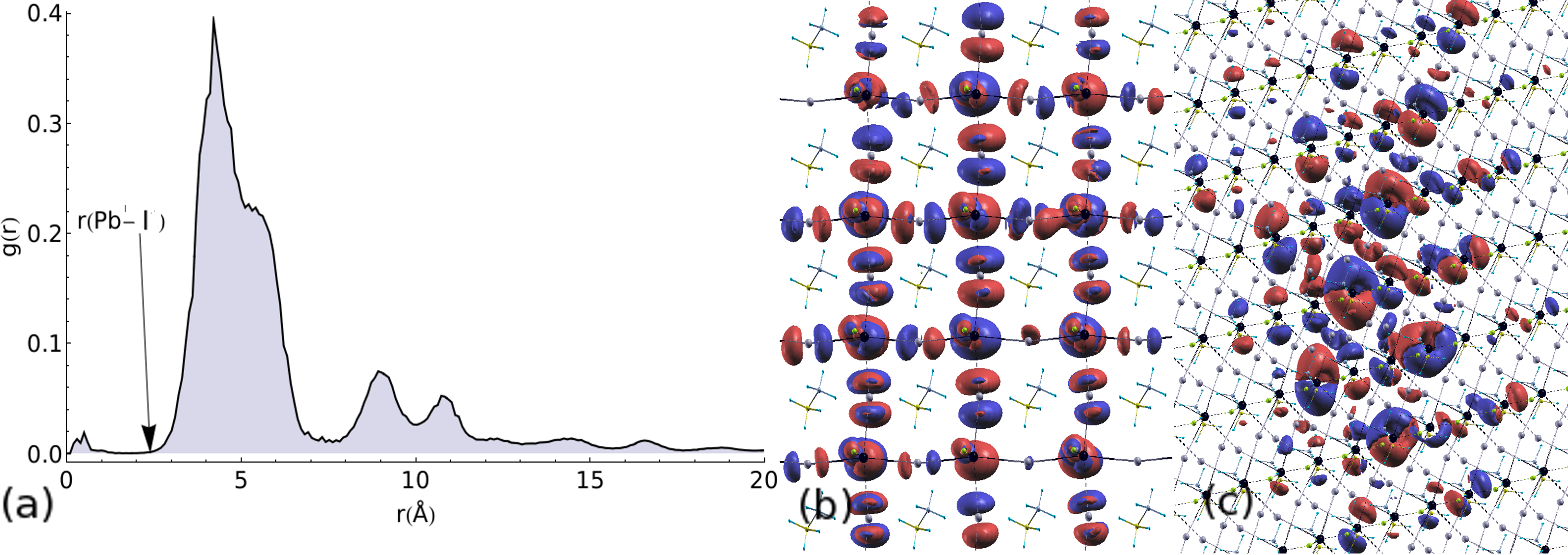}
\caption{(a) The radial distribution function of an electron given that a hole has been placed at the center of a 10x10x10 supercell shows the significant delocalization of the weakly bound Mott-Wannier exciton over several unit cells of the material. The distribution is for an exction of energy ~2.81 eV. (b) The hole density for the same exciton state shows I p-orbital and some Pb s-orbital character of the hole density. (c) The electron density of the same exciton shows Pb p-orbital character. These pictures characterize the excitation as essentially a charge transfer from the iodine to lead. The oscillation in phase between adjacent unit cells is also evident. We verified that all low-energy excitons have similar charge-transfer character.}
\label{fig:fig2}
\end{figure}
%
\textit{Excited state dynamics and AIMD.} \indent Recently, ultrafast transient absorption (TA) experiments have been published examining the tri-iodide perovskite on timescales between 5ps and a nanosecond\cite{TransientAbsorption}. 
This experiment suggests that electrons in the conduction band resist interband relaxation for many nanoseconds, enough to maintain a significant population of electrons and shift absorption. However, there are also relatively fast intraband decay channels visible in the dynamics, especially a rapid redshift of an induced absorption occurring on a timescale on the order of tens of picoseconds. On the basis of our band structure and cluster model studies, the simplified three-band picture of the transient absorption proposed in the TA work is a useful simplification of the genuine excitons of the G0W0-BSE which retain character very similar to the free Kohn-Sham bands. The lowest exciton, $|1\rangle$, is especially localized in k-space, with 95\% of its population coming from the $|\psi_\text{H} \psi_\text{L}\rangle$ transition at R-point, and would correspond in energy to the transition at 760nm observed in the experiment. The transition at 480nm corresponds to excitons which are 0.93eV-1.45eV higher in energy, assuming that its relatively small energetic width indicates it has bound character. We scanned the many exciton states in the energy range corresponding to the 480nm transition and examine two of the lowest excitons, $|2\rangle$ and $|3\rangle$ which are above the lowest by 0.93eV-1.45eV. The electron and hole densities for one of these higher energy excitons, the seventy-fifth exciton with energy 3.37eV which we have labelled as $|2\rangle$ here, is shown in the Supplmentary Information Figures SF4a and SF4b. 

\indent These higher excitons are also strongly localized at the R-point with 89\% and 41\% of their weight coming from the $\alpha |\psi_\text{H} \psi_\text{L+1}\rangle + \beta |\psi_\text{H} \psi_\text{L+2}\rangle$ and $|\psi_\text{H-1} \psi_\text{L}\rangle$ transitions at R-point, respectively. For this reason we will make a simplifying assumption that the phonon dispersion is constant for the purposes of understanding the density of states causing relaxation between these bands. Relaxing this assumption, incorporating the effects of spin-orbit coupling and integrating the dynamics associated with these rates will be topics of future study. It is likely in the TA experiment that 480nm light stimulates these higher-energy transition states, $|2\rangle$ and $|3\rangle$, and the picosecond dynamics seen in the experiment are the relaxation of the higher valence and conduction bands, $|\psi_\text{H-1}\rangle$, $|\psi_\text{L+1}\rangle$, $|\psi_\text{L+2}\rangle$, into the lowest valence and conduction band, $|\psi_\text{H}\rangle$ and $|\psi_\text{L}\rangle$, by emission of vibrational energy to produce the lowest energy electron-hole state $|1\rangle \approx |\psi_\text{H} \psi_\text{L}\rangle$ (the lowest exciton). 
The energy level scheme of our picture of two transitions is shown in Figure 3. Pauli-blocking of electrons in the conduction band is then responsible for the induced bleach of $|1\rangle$, which appears immediately near 760nm.\\

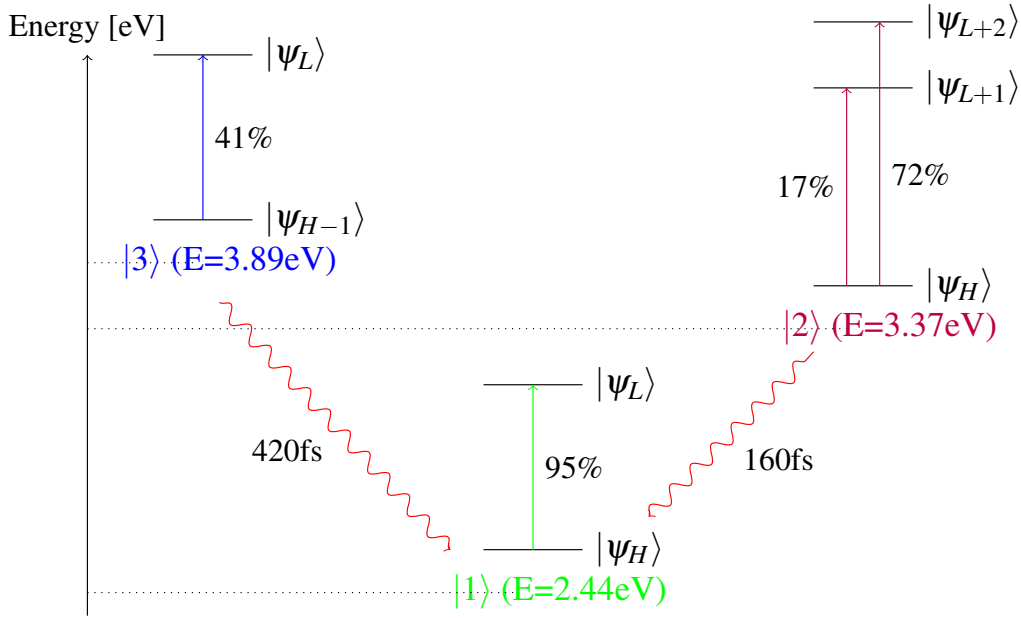
\begin{figure}
\centerline{
  \resizebox{14cm}{!}{
    \begin{tikzpicture}[scale=0.40]
        \draw[->] (-12,-2) -- (-12,15) node[rotate=0,above,yshift=0mm] {\small Energy [eV]} ;
    \begin{scope}
        \draw (-10,10)  --  ++ (3,0) node[right] {$|\psi_{H-1} \rangle$}
         node[below,blue,xshift=-0.28cm,yshift=-0.15cm]{$|3\rangle$ (E=3.89eV)}; 
        \draw (-10,15) --  ++ (3,0) node[right] {$|\psi_{L} \rangle$};
        \draw[dotted] (-12,8.7) --  ++ (3.3,0);
    \end{scope}
	\begin{scope}
        \draw (0,0)  --  ++ (3,0) node[right] {$|\psi_{H} \rangle$} 
          node[below,green,xshift=-0.28cm,yshift=-0.15cm]{$|1\rangle$ (E=2.44eV)}; 
        \draw (0,5) --  ++ (3,0) node[right] {$|\psi_{L} \rangle$};
        \draw[dotted] (-12,-1.3) --  ++ (13,0);
    \end{scope}
	\begin{scope}
        \draw (10,8)  --  ++ (3,0) node[right] {$|\psi_{H} \rangle$}  
          node[below,purple,xshift=-0.28cm,yshift=-0.15cm]{$|2\rangle$ (E=3.37eV)};
        \draw (10,14) --  ++ (3,0) node[right] {$|\psi_{L+1} \rangle$};
        \draw (10,16) --  ++ (3,0) node[right] {$|\psi_{L+2} \rangle$};
        \draw[dotted] (-12,6.7) --  ++ (23,0);
    \end{scope}
    \draw[->,color=blue] (-8.5,10) -- (-8.5,15) node[midway,black,right,font=\small]{41\%};
    \draw[->,color=green] (1.5,0) -- (1.5,5) node[midway,black,right,font=\small]{95\%};
    \begin{scope}[color=purple] 
        \draw[->] (11,8) -- (11,14) node[midway,black,left,font=\small]{17\%};
        \draw[->] (12,8) -- (12,16) node[midway,black,right,yshift=-2.5mm,font=\small]{72\%};
	\end{scope}
    \begin{scope}[color=red]
        \draw[->,decorate, decoration=snake] (-8,7.5) -- (-1,0) node[midway,black,left,font=\small,yshift=-0.3cm]{420fs};
        \draw[<-,decorate, decoration=snake] (5,1) -- (10,6) node[midway,black,right,font=\small,yshift=-0.3cm]{160fs};
    \end{scope}
    \end{tikzpicture}
    }
}
\caption{Energy level diagram showing the dominant KS transitions and their weights determined by G0W0-BSE near the R-point contributing to the three excitons, \ket{1}, \ket{2}, and \ket{3} considered in our model of non-radiative relaxation, and the second order rates for \ket{3},\ket{2}$\rightarrow$ \ket{1} relaxation channels.}
\label{fig:fig3}
\end{figure}

\indent To calculate the intraband relaxation rates, we built a quantum kinetic model using the Redfield rates\cite{anitzan} transitions between states. For each of the Kohn-Sham bands, we extracted the fluctuations in the band energy at k-points within $k \approx 0.1\frac{2\pi}{a}$ of R (a=11.2 Bohr is the lattice constant) from a 5ps of \emph{ab-initio} molecular dynamics simulation at room-temperature (300K), allowing us to study the effect of thermal vibrations on the electronic states of the system.  From these dynamics, we made a semi-classical spectral density (SD)\cite{stephanie} for each Kohn-Sham particle hole state, $J_{c,v,k}(\omega)$, at k-points near R. The SDs obtained from the AIMD are indicative of the strength of the frequency-dependent electronic coupling to the phonon thermal bath. We calculate the SDs for those transitions which contribute most weight to the the excitons $|1\rangle$, $|2\rangle$, and $|3\rangle$, namely, $|\psi_\text{H} \psi_\text{L}\rangle$, $|\psi_\text{H} \psi_\text{L+1}\rangle$, $\beta |\psi_\text{H} \psi_\text{L+2}\rangle$, and $|\psi_\text{H-1} \psi_\text{L}\rangle$. We thus obtain 29 Kohn-Sham transition basis states (channels) labelled by the $c,v,k$ given above. The BSE provides mixing coefficients $A^i_{cvk}$ for each Kohn-Sham transition ($c,v,k$) in the exciton ($|i\rangle$). Based on this mixing and the SDs for each of the KS transitions, we can produce estimates of the relaxation rates (see Table 1) between excitons from a master equation\cite{Parkhill:2012uq,anitzan}: 
\begin{align}
1/\Gamma_{|1\rangle \rightarrow |2\rangle} \propto \sum_{c,v,k} |A^1_{c,v,k}A^2_{c,v,k}|^2 \int_0^\infty \int_{-\infty}^{\infty} J_{c,v,k}(\omega')(e^{i\omega't}n(\omega')+e^{-i\omega't}(n(\omega')+1)   e^{i\omega_{12}t} dtd\omega'
\end{align}
where $n(\omega)$ is the thermal population of a boson at 300K. Once again, this expression assumes an essentially constant phonon dispersion, and ignores the conservation of momentum during an electron phonon scattering event. We are thus confining electronic relaxation to a fixed point in k-space, which here is the R point.  This is a scenario where both hot-electrons and hot-holes in the higher-energy excited states cool by relaxing down to the gap-state which has a lifetime of hundereds of nanoseconds (see Table 1). In addition we also find a faster process of $|2\rangle \rightarrow |1\rangle$ hidden by the former, slower relaxation. It is encouraging to find that our results agree with the experimental values\cite{TransientAbsorption,Xing18102013} of the methylammonium tri-iodide perovskite to the same order of magnitude, however although in our semiclassical treatment this relaxation possesses is extremely efficient, it posesses a degree of semi-classical ambiguity. As discussed in a recent paper, there are a few possible assumptions used to derive $J(\omega)$ leading to slightly different rates. We present the rates as calculated within both the 'harmonic' and 'standard' models for $J(\omega)$. The harmonic model tends to predict larger rates, especially for high frequency bath modes, however both models make the same qualitative predictions. \\

\begin{figure}
\centering
\includegraphics[width=0.5\textwidth]{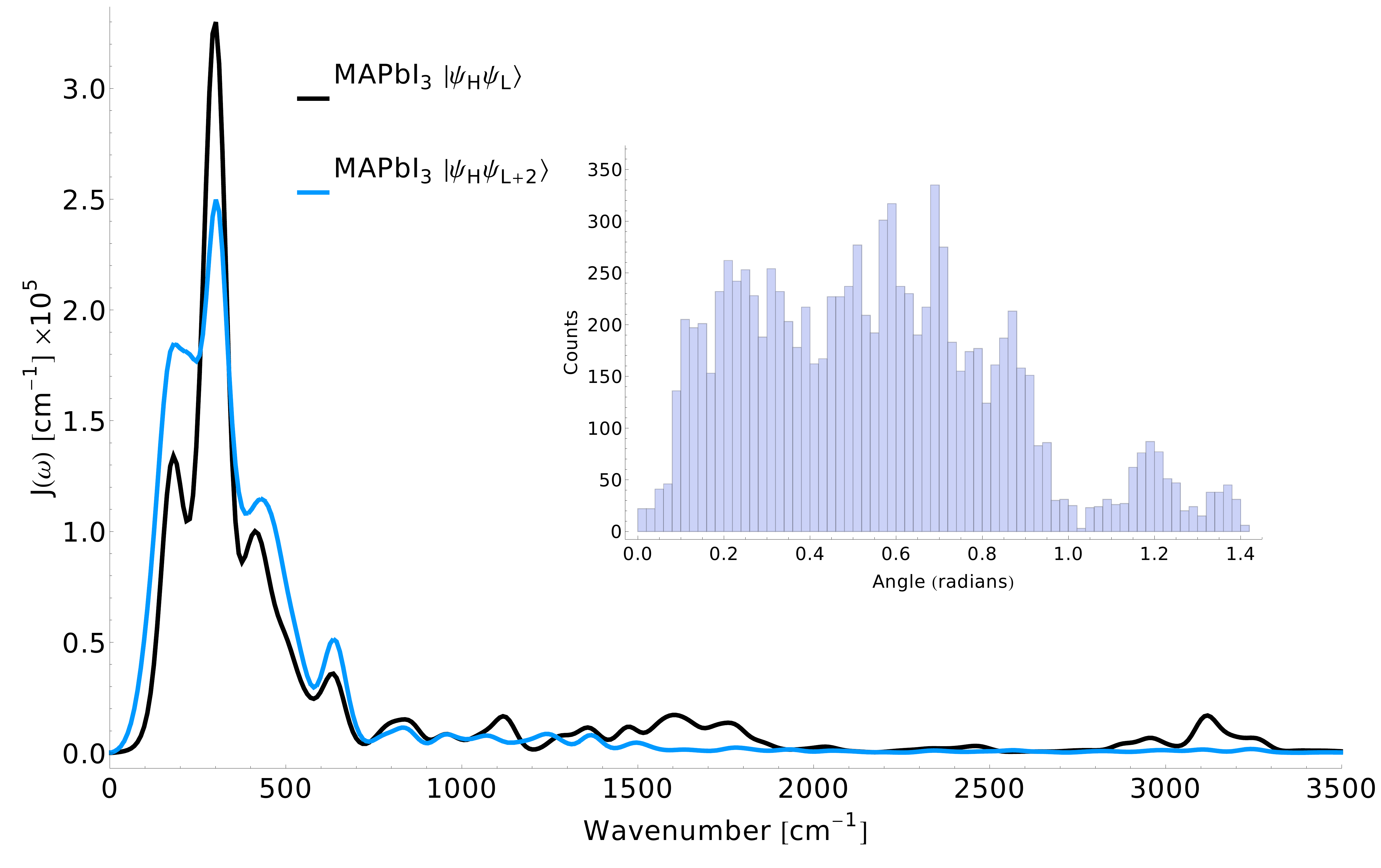}
\caption{The spectral densities, $J(\omega)$, for the KS electron-hole product states $|1\rangle \approx |\psi_\text{H} \psi_\text{L+2}\rangle$ and $|2\rangle \approx |\psi_\text{H} \psi_\text{L}\rangle$ Inset: Histograms of the angular deviations of the methylammonium moiety from its equilibrium orientation in the AIMD for the tri-iodide MAPbI$_3$.}
\label{fig:fig4}
\end{figure}

\indent We now describe the dominant atomic and molecular motions seen in the AIMD. In the room temperature tetragonal phase, the methylammonium moiety is known to be disordered, and thought to rotate with a moderate energetic barrier\cite{Frost:2014aa}. This rotation has been associated with very long timescale ($\sim 1$s) relaxation dynamics \cite{Sanchez:0aa,doi:10.1021/jz501392m}. As expected, the lighter methylammonium moiety dominates the dynamics, followed by the I, and Pb in that order. Most of the motion is rotation along the axis of the methylammonium dipoles; the C-N axis only fluctuates slightly at room temperature on picosecond timescales (Figure 4).  At the R-point, the gap's position, both of the frontier bands couple to the bath more weakly than the bands above. There is greater methylammonium character in the latter electronic states, which even exhibit some coupling to the 3000cm$^{-1}$ C-H vibrations. A comparison of the SDs for our excitons'  dominant particle-hole states around HOMO and LUMO in Figure 4 shows a structure which mirrors the experimental Raman spectrum in Figure 1a of Ref. \onlinecite{Quarti_raman_spectrum_2013} between 0-500 cm$^{-1}$. The bulk of the relaxation rate comes from rocking and spining motion of CH$_3$NH$_3$ along it's long axis.\\

\textit{Conclusion.} \indent In this paper, we computed the excited states and optical absorption spectrum of the tri-iodide perovskite CH$_{3}$NH$_{3}$PbI$_{3}$ using DFT+GW+BSE. We also built a model of the electronic system coupled to the system's phonons at $\Gamma$ point to describe non-radiative relaxation between these excited states, using data from 5ps of AIMD. Specifically, we computed the rates of relaxation using the Redfield equation-of-motion for the reduced density matrix of the electronic system, using spectral densities from AIMD to describe the coupling between the electronic system and phonon bath. We found good agreement between our rates and the experimental rates for the methylammonium tri-iodide, as well as agreement between the calculated and experimental optical absorption spectrum for the tri-iodide perovskite.

\begin{table}
\centering
\caption{Relaxation rates for intraband transitions in chloride tri-iodide lead perovskites calculated using Eqn. 1 on the basis of a three exciton model, where the three excitons have the dominant transitions: $|3\rangle \approx |\psi_\text{H-1} \psi_\text{L}\rangle$, $|1\rangle \approx |\psi_\text{H} \psi_\text{L}\rangle$, $|2\rangle \approx \alpha |\psi_\text{H} \psi_\text{L+1}\rangle + \beta |\psi_\text{H} \psi_\text{L+2}\rangle$. The difference between the Standard and harmonic semiclassical model is in the braces. E$_1$=2.44eV is the energy of the lowest singlet exciton.}
\begin{tabular}{| c | c | c | c | c | c |}
  \hline                       
  Halide    & Standard (Harm.) & Energy (eV) & Standard (Harm.) & Energy (eV) & Expt. (fs)  \\
            &  $|3\rangle \rightarrow |1\rangle$ (fs) & $|3\rangle$ & $|2\rangle \rightarrow |1\rangle$ (fs) & $|2\rangle$ &            \\ \hline
  PbI$_3$   & 420 (-100) & - & 160 (-39.8) & - & 400 \\ \hline
\end{tabular}
\label{tab:table1}
\end{table}


\acknowledgement
We thank The University of Notre Dame's College of Science and Department of Chemistry and Biochemistry, NDEnergy and the Honeywell Corporation for generous start-up funding. This research used resources of the National Energy Research Scientific Computing Center, which is supported by the Office of Science of the U.S. Department of Energy under Contract No. DE-AC02-05CH11231.

\section{Associated content}
Supporting information includes details of computational input parameters for DFT and GW+BSE calculations, table of structure optimization, plots of bandstructure and density-of-states, natural transition orbitals of the first excited state from TD-DFT cluster calculations, electron and hole densities of exciton $|2\rangle$ with energy 3.37eV. This material is available free of charge via the Internet at http://pubs.acs.org

\section{Author information}
Corresponding author:
John Parkhill, email: jparkhil@nd.edu

\section{Notes}
The authors declare no competing financial interest.

\bibliography{pskite}

\providecommand{\latin}[1]{#1}
\providecommand*\mcitethebibliography{\thebibliography}
\csname @ifundefined\endcsname{endmcitethebibliography}
  {\let\endmcitethebibliography\endthebibliography}{}
\begin{mcitethebibliography}{43}
\providecommand*\natexlab[1]{#1}
\providecommand*\mciteSetBstSublistMode[1]{}
\providecommand*\mciteSetBstMaxWidthForm[2]{}
\providecommand*\mciteBstWouldAddEndPuncttrue
  {\def\EndOfBibitem{\unskip.}}
\providecommand*\mciteBstWouldAddEndPunctfalse
  {\let\EndOfBibitem\relax}
\providecommand*\mciteSetBstMidEndSepPunct[3]{}
\providecommand*\mciteSetBstSublistLabelBeginEnd[3]{}
\providecommand*\EndOfBibitem{}
\mciteSetBstSublistMode{f}
\mciteSetBstMaxWidthForm{subitem}{(\alph{mcitesubitemcount})}
\mciteSetBstSublistLabelBeginEnd
  {\mcitemaxwidthsubitemform\space}
  {\relax}
  {\relax}

\bibitem[Snaith(2013)]{snaith_perovskites_2013}
Snaith,~H.~J. \emph{The Journal of Physical Chemistry Letters} \textbf{2013},
  3623--3630\relax
\mciteBstWouldAddEndPuncttrue
\mciteSetBstMidEndSepPunct{\mcitedefaultmidpunct}
{\mcitedefaultendpunct}{\mcitedefaultseppunct}\relax
\EndOfBibitem
\bibitem[Ball \latin{et~al.}(2013)Ball, Lee, Hey, and
  Snaith]{snaith_EESci_2013}
Ball,~J.~M.; Lee,~M.~M.; Hey,~A.; Snaith,~H.~J. \emph{Energy Environ. Sci.}
  \textbf{2013}, \emph{6}, 1739--1743\relax
\mciteBstWouldAddEndPuncttrue
\mciteSetBstMidEndSepPunct{\mcitedefaultmidpunct}
{\mcitedefaultendpunct}{\mcitedefaultseppunct}\relax
\EndOfBibitem
\bibitem[Liu \latin{et~al.}(2013)Liu, Johnston, and Snaith]{liu_efficient_2013}
Liu,~M.; Johnston,~M.~B.; Snaith,~H.~J. \emph{Nature} \textbf{2013},
  \emph{501}, 395--398\relax
\mciteBstWouldAddEndPuncttrue
\mciteSetBstMidEndSepPunct{\mcitedefaultmidpunct}
{\mcitedefaultendpunct}{\mcitedefaultseppunct}\relax
\EndOfBibitem
\bibitem[Malinkiewicz \latin{et~al.}(2014)Malinkiewicz, Yella, Lee,
  Espallargas, Graetzel, Nazeeruddin, and Bolink]{Malinkiewicz}
Malinkiewicz,~O.; Yella,~A.; Lee,~Y.~H.; Espallargas,~G.~M.; Graetzel,~M.;
  Nazeeruddin,~M.~K.; Bolink,~H.~J. \emph{Nat Photon} \textbf{2014}, \emph{8},
  128--132\relax
\mciteBstWouldAddEndPuncttrue
\mciteSetBstMidEndSepPunct{\mcitedefaultmidpunct}
{\mcitedefaultendpunct}{\mcitedefaultseppunct}\relax
\EndOfBibitem
\bibitem[Liu and Kelly(2014)Liu, and Kelly]{Kelly}
Liu,~D.; Kelly,~T.~L. \emph{Nat Photon} \textbf{2014}, \emph{8}, 133--138\relax
\mciteBstWouldAddEndPuncttrue
\mciteSetBstMidEndSepPunct{\mcitedefaultmidpunct}
{\mcitedefaultendpunct}{\mcitedefaultseppunct}\relax
\EndOfBibitem
\bibitem[Bretschneider \latin{et~al.}(2014)Bretschneider, Weickert, Dorman, and
  Schmidt-Mende]{Bretschneider:2014aa}
Bretschneider,~S.~A.; Weickert,~J.; Dorman,~J.~A.; Schmidt-Mende,~L. \emph{APL
  Materials} \textbf{2014}, \emph{2}, 040701\relax
\mciteBstWouldAddEndPuncttrue
\mciteSetBstMidEndSepPunct{\mcitedefaultmidpunct}
{\mcitedefaultendpunct}{\mcitedefaultseppunct}\relax
\EndOfBibitem
\bibitem[Edri \latin{et~al.}(2013)Edri, Kirmayer, Cahen, and
  Hodes]{Edri:2013aa}
Edri,~E.; Kirmayer,~S.; Cahen,~D.; Hodes,~G. \emph{The Journal of Physical
  Chemistry Letters} \textbf{2013}, \emph{4}, 897--902\relax
\mciteBstWouldAddEndPuncttrue
\mciteSetBstMidEndSepPunct{\mcitedefaultmidpunct}
{\mcitedefaultendpunct}{\mcitedefaultseppunct}\relax
\EndOfBibitem
\bibitem[Baikie \latin{et~al.}(2013)Baikie, Fang, Kadro, Schreyer, Wei,
  Mhaisalkar, Graetzel, and White]{Baikie:2013aa}
Baikie,~T.; Fang,~Y.; Kadro,~J.~M.; Schreyer,~M.; Wei,~F.; Mhaisalkar,~S.~G.;
  Graetzel,~M.; White,~T.~J. \emph{J. Mater. Chem. A} \textbf{2013}, \emph{1},
  5628--5641\relax
\mciteBstWouldAddEndPuncttrue
\mciteSetBstMidEndSepPunct{\mcitedefaultmidpunct}
{\mcitedefaultendpunct}{\mcitedefaultseppunct}\relax
\EndOfBibitem
\bibitem[Christians \latin{et~al.}(2014)Christians, Fung, and
  Kamat]{doi:10.1021/ja411014k}
Christians,~J.~A.; Fung,~R. C.~M.; Kamat,~P.~V. \emph{Journal of the American
  Chemical Society} \textbf{2014}, \emph{136}, 758--764\relax
\mciteBstWouldAddEndPuncttrue
\mciteSetBstMidEndSepPunct{\mcitedefaultmidpunct}
{\mcitedefaultendpunct}{\mcitedefaultseppunct}\relax
\EndOfBibitem
\bibitem[Hao \latin{et~al.}(2014)Hao, Stoumpos, Cao, Chang, and
  Kanatzidis]{leadfree}
Hao,~F.; Stoumpos,~C.~C.; Cao,~D.~H.; Chang,~R. P.~H.; Kanatzidis,~M.~G.
  \emph{Nat Photon} \textbf{2014}, \emph{8}, 489--494\relax
\mciteBstWouldAddEndPuncttrue
\mciteSetBstMidEndSepPunct{\mcitedefaultmidpunct}
{\mcitedefaultendpunct}{\mcitedefaultseppunct}\relax
\EndOfBibitem
\bibitem[Green \latin{et~al.}(2014)Green, Ho-Baillie, and Snaith]{Green}
Green,~M.~A.; Ho-Baillie,~A.; Snaith,~H.~J. \emph{Nat Photon} \textbf{2014},
  \emph{8}, 506--514\relax
\mciteBstWouldAddEndPuncttrue
\mciteSetBstMidEndSepPunct{\mcitedefaultmidpunct}
{\mcitedefaultendpunct}{\mcitedefaultseppunct}\relax
\EndOfBibitem
\bibitem[Wehrenfennig \latin{et~al.}(2014)Wehrenfennig, Eperon, Johnston,
  Snaith, and Herz]{mob_snaith}
Wehrenfennig,~C.; Eperon,~G.~E.; Johnston,~M.~B.; Snaith,~H.~J.; Herz,~L.~M.
  \emph{Advanced Materials} \textbf{2014}, \emph{26}, 1584--1589\relax
\mciteBstWouldAddEndPuncttrue
\mciteSetBstMidEndSepPunct{\mcitedefaultmidpunct}
{\mcitedefaultendpunct}{\mcitedefaultseppunct}\relax
\EndOfBibitem
\bibitem[Xing \latin{et~al.}(2013)Xing, Mathews, Sun, Lim, Lam, Grätzel,
  Mhaisalkar, and Sum]{Xing18102013}
Xing,~G.; Mathews,~N.; Sun,~S.; Lim,~S.~S.; Lam,~Y.~M.; Grätzel,~M.;
  Mhaisalkar,~S.; Sum,~T.~C. \emph{Science} \textbf{2013}, \emph{342},
  344--347\relax
\mciteBstWouldAddEndPuncttrue
\mciteSetBstMidEndSepPunct{\mcitedefaultmidpunct}
{\mcitedefaultendpunct}{\mcitedefaultseppunct}\relax
\EndOfBibitem
\bibitem[Du(2014)]{C4TA01198H}
Du,~M.~H. \emph{J. Mater. Chem. A} \textbf{2014}, \emph{2}, 9091--9098\relax
\mciteBstWouldAddEndPuncttrue
\mciteSetBstMidEndSepPunct{\mcitedefaultmidpunct}
{\mcitedefaultendpunct}{\mcitedefaultseppunct}\relax
\EndOfBibitem
\bibitem[Savenije \latin{et~al.}(2014)Savenije, Ponseca, Kunneman, Abdellah,
  Zheng, Tian, Zhu, Canton, Scheblykin, Pullerits, Yartsev, and
  Sundstr{\"o}m]{Savenije:0aa}
Savenije,~T.~J.; Ponseca,~C.~S.; Kunneman,~L.; Abdellah,~M.; Zheng,~K.;
  Tian,~Y.; Zhu,~Q.; Canton,~S.~E.; Scheblykin,~I.~G.; Pullerits,~T.;
  Yartsev,~A.; Sundstr{\"o}m,~V. \emph{The Journal of Physical Chemistry
  Letters} \textbf{2014}, \emph{5}, 2189--2194\relax
\mciteBstWouldAddEndPuncttrue
\mciteSetBstMidEndSepPunct{\mcitedefaultmidpunct}
{\mcitedefaultendpunct}{\mcitedefaultseppunct}\relax
\EndOfBibitem
\bibitem[Zhang \latin{et~al.}(2013)Zhang, Saliba, Stranks, Sun, Shi, Wiesner,
  and Snaith]{Zhang:2013aa}
Zhang,~W.; Saliba,~M.; Stranks,~S.~D.; Sun,~Y.; Shi,~X.; Wiesner,~U.;
  Snaith,~H.~J. \emph{Nano Letters} \textbf{2013}, \emph{13}, 4505--4510\relax
\mciteBstWouldAddEndPuncttrue
\mciteSetBstMidEndSepPunct{\mcitedefaultmidpunct}
{\mcitedefaultendpunct}{\mcitedefaultseppunct}\relax
\EndOfBibitem
\bibitem[DaInnocenzo \latin{et~al.}(2014)DaInnocenzo, Grancini, Alcocer,
  Kandada, Stranks, Lee, Lanzani, Snaith, and Petrozza]{DaInnocenzo:2014aa}
DaInnocenzo,~V.; Grancini,~G.; Alcocer,~M. J.~P.; Kandada,~A. R.~S.;
  Stranks,~S.~D.; Lee,~M.~M.; Lanzani,~G.; Snaith,~H.~J.; Petrozza,~A.
  \emph{Nat Commun} \textbf{2014}, \emph{5}\relax
\mciteBstWouldAddEndPuncttrue
\mciteSetBstMidEndSepPunct{\mcitedefaultmidpunct}
{\mcitedefaultendpunct}{\mcitedefaultseppunct}\relax
\EndOfBibitem
\bibitem[Brivio \latin{et~al.}(2014)Brivio, Butler, Walsh, and van
  Schilfgaarde]{PhysRevB.89.155204}
Brivio,~F.; Butler,~K.~T.; Walsh,~A.; van Schilfgaarde,~M. \emph{Phys. Rev. B}
  \textbf{2014}, \emph{89}, 155204\relax
\mciteBstWouldAddEndPuncttrue
\mciteSetBstMidEndSepPunct{\mcitedefaultmidpunct}
{\mcitedefaultendpunct}{\mcitedefaultseppunct}\relax
\EndOfBibitem
\bibitem[Umari \latin{et~al.}(2014)Umari, Mosconi, and
  De~Angelis]{Umari:2014aa}
Umari,~P.; Mosconi,~E.; De~Angelis,~F. \emph{Sci. Rep.} \textbf{2014},
  \emph{4}\relax
\mciteBstWouldAddEndPuncttrue
\mciteSetBstMidEndSepPunct{\mcitedefaultmidpunct}
{\mcitedefaultendpunct}{\mcitedefaultseppunct}\relax
\EndOfBibitem
\bibitem[Zhu \latin{et~al.}(0)Zhu, Su, Marcus, and Michel-Beyerle]{marcus}
Zhu,~X.; Su,~H.; Marcus,~R.~A.; Michel-Beyerle,~M.~E. \emph{The Journal of
  Physical Chemistry Letters} \textbf{0}, \emph{0}, 3061--3065\relax
\mciteBstWouldAddEndPuncttrue
\mciteSetBstMidEndSepPunct{\mcitedefaultmidpunct}
{\mcitedefaultendpunct}{\mcitedefaultseppunct}\relax
\EndOfBibitem
\bibitem[Zhang \latin{et~al.}(2014)Zhang, Yu, Lyu, Wang, Yun, and
  Wang]{C4CC04973J}
Zhang,~M.; Yu,~H.; Lyu,~M.; Wang,~Q.; Yun,~J.-H.; Wang,~L. \emph{Chem. Commun.}
  \textbf{2014}, \emph{50}, 11727--11730\relax
\mciteBstWouldAddEndPuncttrue
\mciteSetBstMidEndSepPunct{\mcitedefaultmidpunct}
{\mcitedefaultendpunct}{\mcitedefaultseppunct}\relax
\EndOfBibitem
\bibitem[Gratzel(2014)]{MGratzel2014}
Gratzel,~M. \emph{Nature Materials} \textbf{2014}, \emph{13}, 838--842\relax
\mciteBstWouldAddEndPuncttrue
\mciteSetBstMidEndSepPunct{\mcitedefaultmidpunct}
{\mcitedefaultendpunct}{\mcitedefaultseppunct}\relax
\EndOfBibitem
\bibitem[Giannozzi \latin{et~al.}(2009)Giannozzi, Baroni, Bonini, Calandra,
  Car, Cavazzoni, Ceresoli, Chiarotti, Cococcioni, Dabo, {Dal Corso},
  de~Gironcoli, Fabris, Fratesi, Gebauer, Gerstmann, Gougoussis, Kokalj,
  Lazzeri, Martin-Samos, Marzari, Mauri, Mazzarello, Paolini, Pasquarello,
  Paulatto, Sbraccia, Scandolo, Sclauzero, Seitsonen, Smogunov, Umari, and
  Wentzcovitch]{QE-2009}
Giannozzi,~P. \latin{et~al.}  \emph{Journal of Physics: Condensed Matter}
  \textbf{2009}, \emph{21}, 395502 (19pp)\relax
\mciteBstWouldAddEndPuncttrue
\mciteSetBstMidEndSepPunct{\mcitedefaultmidpunct}
{\mcitedefaultendpunct}{\mcitedefaultseppunct}\relax
\EndOfBibitem
\bibitem[Troullier and Martins(1991)Troullier, and Martins]{MTpseudo}
Troullier,~N.; Martins,~J.~L. \emph{Phys. Rev. B} \textbf{1991}, \emph{43},
  1993--2006\relax
\mciteBstWouldAddEndPuncttrue
\mciteSetBstMidEndSepPunct{\mcitedefaultmidpunct}
{\mcitedefaultendpunct}{\mcitedefaultseppunct}\relax
\EndOfBibitem
\bibitem[Mosconi \latin{et~al.}(2013)Mosconi, Amat, Nazeeruddin, Gratzel, and
  Angelis]{Mosconi_first-principles-no-soc_2013}
Mosconi,~E.; Amat,~A.; Nazeeruddin,~M.~K.; Gratzel,~M.; Angelis,~F.~D. \emph{J.
  Phys. Chem. C} \textbf{2013}, \emph{117}, 13902−13913\relax
\mciteBstWouldAddEndPuncttrue
\mciteSetBstMidEndSepPunct{\mcitedefaultmidpunct}
{\mcitedefaultendpunct}{\mcitedefaultseppunct}\relax
\EndOfBibitem
\bibitem[Jacky~Even and Katan(2013)Jacky~Even, and Katan]{Even_SOC_calc_2013}
Jacky~Even,~J.-M.~J.,~Laurent~Pedesseau; Katan,~C. \emph{The Journal of
  Physical Chemistry Letters} \textbf{2013}, \emph{4}, 2999--3005\relax
\mciteBstWouldAddEndPuncttrue
\mciteSetBstMidEndSepPunct{\mcitedefaultmidpunct}
{\mcitedefaultendpunct}{\mcitedefaultseppunct}\relax
\EndOfBibitem
\bibitem[Lee \latin{et~al.}(2012)Lee, Teuscher, Miyasaka, Murakami, and
  Snaith]{Lee02112012}
Lee,~M.~M.; Teuscher,~J.; Miyasaka,~T.; Murakami,~T.~N.; Snaith,~H.~J.
  \emph{Science} \textbf{2012}, \emph{338}, 643--647\relax
\mciteBstWouldAddEndPuncttrue
\mciteSetBstMidEndSepPunct{\mcitedefaultmidpunct}
{\mcitedefaultendpunct}{\mcitedefaultseppunct}\relax
\EndOfBibitem
\bibitem[Hedin(1965)]{PhysRev.139.A796}
Hedin,~L. \emph{Phys. Rev.} \textbf{1965}, \emph{139}, A796--A823\relax
\mciteBstWouldAddEndPuncttrue
\mciteSetBstMidEndSepPunct{\mcitedefaultmidpunct}
{\mcitedefaultendpunct}{\mcitedefaultseppunct}\relax
\EndOfBibitem
\bibitem[Hybertsen and Louie(1986)Hybertsen, and Louie]{PhysRevB.34.5390}
Hybertsen,~M.~S.; Louie,~S.~G. \emph{Phys. Rev. B} \textbf{1986}, \emph{34},
  5390--5413\relax
\mciteBstWouldAddEndPuncttrue
\mciteSetBstMidEndSepPunct{\mcitedefaultmidpunct}
{\mcitedefaultendpunct}{\mcitedefaultseppunct}\relax
\EndOfBibitem
\bibitem[Rohlfing and Louie(2000)Rohlfing, and Louie]{PhysRevB.62.4927}
Rohlfing,~M.; Louie,~S.~G. \emph{Phys. Rev. B} \textbf{2000}, \emph{62},
  4927--4944\relax
\mciteBstWouldAddEndPuncttrue
\mciteSetBstMidEndSepPunct{\mcitedefaultmidpunct}
{\mcitedefaultendpunct}{\mcitedefaultseppunct}\relax
\EndOfBibitem
\bibitem[Deslippe \latin{et~al.}(2012)Deslippe, Samsonidze, Strubbe, Jain,
  Cohen, and Louie]{Deslippe20121269}
Deslippe,~J.; Samsonidze,~G.; Strubbe,~D.~A.; Jain,~M.; Cohen,~M.~L.;
  Louie,~S.~G. \emph{Computer Physics Communications} \textbf{2012},
  \emph{183}, 1269 -- 1289\relax
\mciteBstWouldAddEndPuncttrue
\mciteSetBstMidEndSepPunct{\mcitedefaultmidpunct}
{\mcitedefaultendpunct}{\mcitedefaultseppunct}\relax
\EndOfBibitem
\bibitem[{D’Innocenzo} \latin{et~al.}(2014){D’Innocenzo}, Grancini,
  Alcocer, Kandada, Stranks, Lee, Lanzani, Snaith, and
  Petrozza]{dinnocenzo_excitons_2014}
{D’Innocenzo},~V.; Grancini,~G.; Alcocer,~M. J.~P.; Kandada,~A. R.~S.;
  Stranks,~S.~D.; Lee,~M.~M.; Lanzani,~G.; Snaith,~H.~J.; Petrozza,~A.
  \emph{Nature Communications} \textbf{2014}, \emph{5}\relax
\mciteBstWouldAddEndPuncttrue
\mciteSetBstMidEndSepPunct{\mcitedefaultmidpunct}
{\mcitedefaultendpunct}{\mcitedefaultseppunct}\relax
\EndOfBibitem
\bibitem[Gay(1971)]{PhysRevB.4.2567}
Gay,~J.~G. \emph{Phys. Rev. B} \textbf{1971}, \emph{4}, 2567--2575\relax
\mciteBstWouldAddEndPuncttrue
\mciteSetBstMidEndSepPunct{\mcitedefaultmidpunct}
{\mcitedefaultendpunct}{\mcitedefaultseppunct}\relax
\EndOfBibitem
\bibitem[Shao \latin{et~al.}()Shao, Gan, Epifanovsky, Gilbert, Wormit,
  Kussmann, Lange, Behn, Deng, Feng, Ghosh, Goldey, Horn, Jacobson, Kaliman,
  Khaliullin, Ku{\'s}, Landau, Liu, Proynov, Rhee, Richard, Rohrdanz, Steele,
  Sundstrom, Woodcock, Zimmerman, Zuev, Albrecht, Alguire, Austin, Beran,
  Bernard, Berquist, Brandhorst, Bravaya, Brown, Casanova, Chang, Chen, Chien,
  Closser, Crittenden, Diedenhofen, DiStasio, Do, Dutoi, Edgar, Fatehi,
  Fusti-Molnar, Ghysels, Golubeva-Zadorozhnaya, Gomes, Hanson-Heine, Harbach,
  Hauser, Hohenstein, Holden, Jagau, Ji, Kaduk, Khistyaev, Kim, Kim, King,
  Klunzinger, Kosenkov, Kowalczyk, Krauter, Lao, Laurent, Lawler, Levchenko,
  Lin, Liu, Livshits, Lochan, Luenser, Manohar, Manzer, Mao, Mardirossian,
  Marenich, Maurer, Mayhall, Neuscamman, Oana, Olivares-Amaya, O'Neill,
  Parkhill, Perrine, Peverati, Prociuk, Rehn, Rosta, Russ, Sharada, Sharma,
  Small, Sodt, Stein, St{\"u}ck, Su, Thom, Tsuchimochi, Vanovschi, Vogt,
  Vydrov, Wang, Watson, Wenzel, White, Williams, Yang, Yeganeh, Yost, You,
  Zhang, Zhang, Zhao, Brooks, Chan, Chipman, Cramer, Goddard, Gordon, Hehre,
  Klamt, Schaefer, Schmidt, Sherrill, Truhlar, Warshel, Xu, Aspuru-Guzik, Baer,
  Bell, Besley, Chai, Dreuw, Dunietz, Furlani, Gwaltney, Hsu, Jung, Kong,
  Lambrecht, Liang, Ochsenfeld, Rassolov, Slipchenko, Subotnik, Van~Voorhis,
  Herbert, Krylov, Gill, and Head-Gordon]{QChem2014}
Shao,~Y. \latin{et~al.}  \emph{Molecular Physics} \emph{(advance online)},
  1--32\relax
\mciteBstWouldAddEndPuncttrue
\mciteSetBstMidEndSepPunct{\mcitedefaultmidpunct}
{\mcitedefaultendpunct}{\mcitedefaultseppunct}\relax
\EndOfBibitem
\bibitem[Manser and Kamat(2014)Manser, and Kamat]{TransientAbsorption}
Manser,~J.~S.; Kamat,~P.~V. \emph{Nat Photon} \textbf{2014}, \emph{advance
  online publication}, --\relax
\mciteBstWouldAddEndPuncttrue
\mciteSetBstMidEndSepPunct{\mcitedefaultmidpunct}
{\mcitedefaultendpunct}{\mcitedefaultseppunct}\relax
\EndOfBibitem
\bibitem[Nitzan(2006)]{anitzan}
Nitzan,~A. \emph{Chemical Dynamics in Condensed Phases} \textbf{2006},
  368--389\relax
\mciteBstWouldAddEndPuncttrue
\mciteSetBstMidEndSepPunct{\mcitedefaultmidpunct}
{\mcitedefaultendpunct}{\mcitedefaultseppunct}\relax
\EndOfBibitem
\bibitem[Valleau \latin{et~al.}(2012)Valleau, Eisfeld, and
  Aspuru-Guzik]{stephanie}
Valleau,~S.; Eisfeld,~A.; Aspuru-Guzik,~A. \emph{The Journal of Chemical
  Physics} \textbf{2012}, \emph{137}, 224103\relax
\mciteBstWouldAddEndPuncttrue
\mciteSetBstMidEndSepPunct{\mcitedefaultmidpunct}
{\mcitedefaultendpunct}{\mcitedefaultseppunct}\relax
\EndOfBibitem
\bibitem[Parkhill \latin{et~al.}(2012)Parkhill, Markovich, Tempel, and
  Aspuru-Guzik]{Parkhill:2012uq}
Parkhill,~J.~A.; Markovich,~T.; Tempel,~D.~G.; Aspuru-Guzik,~A. \emph{The
  Journal of Chemical Physics} \textbf{2012}, \emph{137}, 22A547\relax
\mciteBstWouldAddEndPuncttrue
\mciteSetBstMidEndSepPunct{\mcitedefaultmidpunct}
{\mcitedefaultendpunct}{\mcitedefaultseppunct}\relax
\EndOfBibitem
\bibitem[Frost \latin{et~al.}(2014)Frost, Butler, Brivio, Hendon, van
  Schilfgaarde, and Walsh]{Frost:2014aa}
Frost,~J.~M.; Butler,~K.~T.; Brivio,~F.; Hendon,~C.~H.; van Schilfgaarde,~M.;
  Walsh,~A. \emph{Nano Letters} \textbf{2014}, \emph{14}, 2584--2590\relax
\mciteBstWouldAddEndPuncttrue
\mciteSetBstMidEndSepPunct{\mcitedefaultmidpunct}
{\mcitedefaultendpunct}{\mcitedefaultseppunct}\relax
\EndOfBibitem
\bibitem[Sanchez \latin{et~al.}(2014)Sanchez, Gonzalez-Pedro, Lee, Park, Kang,
  Mora-Sero, and Bisquert]{Sanchez:0aa}
Sanchez,~R.~S.; Gonzalez-Pedro,~V.; Lee,~J.-W.; Park,~N.-G.; Kang,~Y.~S.;
  Mora-Sero,~I.; Bisquert,~J. \emph{The Journal of Physical Chemistry Letters}
  \textbf{2014}, \emph{5}, 2357--2363\relax
\mciteBstWouldAddEndPuncttrue
\mciteSetBstMidEndSepPunct{\mcitedefaultmidpunct}
{\mcitedefaultendpunct}{\mcitedefaultseppunct}\relax
\EndOfBibitem
\bibitem[Kim and Park(2014)Kim, and Park]{doi:10.1021/jz501392m}
Kim,~H.-S.; Park,~N.-G. \emph{The Journal of Physical Chemistry Letters}
  \textbf{2014}, \emph{5}, 2927--2934\relax
\mciteBstWouldAddEndPuncttrue
\mciteSetBstMidEndSepPunct{\mcitedefaultmidpunct}
{\mcitedefaultendpunct}{\mcitedefaultseppunct}\relax
\EndOfBibitem
\bibitem[Claudio \latin{et~al.}(2013)Claudio, Grancini, Mosconi, Bruno, Ball,
  Lee, Snaith, Petrozza, and Angelis]{Quarti_raman_spectrum_2013}
Claudio,~Q.; Grancini,~G.; Mosconi,~E.; Bruno,~P.; Ball,~J.~M.; Lee,~M.~M.;
  Snaith,~H.~J.; Petrozza,~A.; Angelis,~F.~D. \emph{J. Phys. Chem. Lett.}
  \textbf{2013}, \emph{5}, 279−284\relax
\mciteBstWouldAddEndPuncttrue
\mciteSetBstMidEndSepPunct{\mcitedefaultmidpunct}
{\mcitedefaultendpunct}{\mcitedefaultseppunct}\relax
\EndOfBibitem
\end{mcitethebibliography}

\providecommand{\latin}[1]{#1}
\providecommand*\mcitethebibliography{\thebibliography}
\csname @ifundefined\endcsname{endmcitethebibliography}
  {\let\endmcitethebibliography\endthebibliography}{}

\end{document}


\section{Computational Methodology}

\indent For the DFT+PBE calculations, the Quantum Espresso package was used with the scalar relativistic MT pseudopotentials. Cutoffs of 25/200 Ry were used to generate the AIMD, while 50/250 Ry were used for the SCF. A 4x4x4 k-point sampling was used for both the AIMD and SCF. For the TDDFT and RI-SOS-ADC(2) cluster calculations, the LANL2dz and RIMP2-pvdz basis and aux-basis sets were used. \\

\indent For the GW+BSE, the following parameters were used: 1) For the epsilon calculation, 25 valence bands, 275 conduction bands, 64 k-points in the coarse grid, and a cutoff of 10 Ry was used; 2) For sigma, the screened and bare Coulomb cut-offs were 10.0 and 25.0 Ry respectively, and the same number of valence and conduction bands were used; 3) For the kernel calculation, the same Coulomb cut-offs as for the sigma were used, and the BSE interaction matrix elements were calculated between 12 valence and 24 conduction bands; and 4) For the absorption calculation, the diagonalization was carried out between the same valence and conduction bands as calculated in the kernel, and a fine grid of 512 k-points was used. \\

\section{Tables}

\begin{table}[]
\centering
\caption{Summary of optimization of crystal structures using Quantum Espresso}
\label{tab:ST1}
\begin{tabular}{| c | c | c | c | c | c | c | c | c |}
  \hline                       
  cell & a (A$^o$) & b (A$^o$) & c (A$^o$) & grid & SOC & gap (eV) & total energy (Ry) & total force     \\ \hline
  primitive & 6.243 & 6.298 & 5.593 & 4x4x4 & no & 2.53 &  -265.07918148  &  0.002520   \\ \hline
  primitive & 6.545 & 6.545 & 6.488 & 4x4x4 & scalar & 2.75 & -234.25295292  & 0.004144 \\ \hline
\end{tabular}
\end{table}

\section{Figures}

\begin{figure}
\centering
\includegraphics[width=0.5\textwidth]{bandstructure.png}
\caption{Band structure with MT pseudopotential, 4x4x4 k-grid, ecut = 50/250 Ry.}
\label{fig:SF1}
\end{figure}

\begin{figure}
\centering
\includegraphics[width=0.5\textwidth]{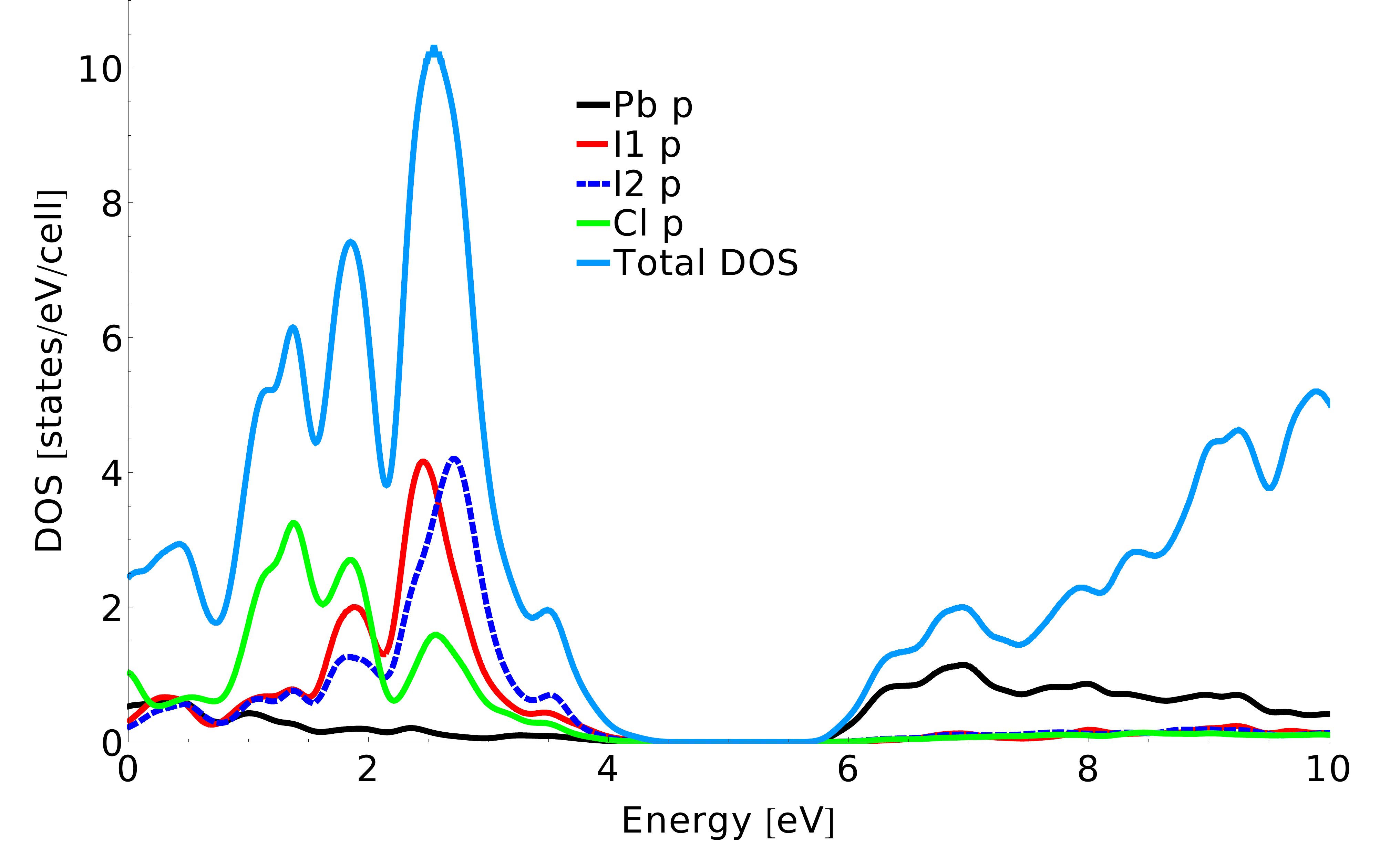}
\caption{DOS without SOC, 4x4x4 k-grid, ecut = 50/250 Ry.}
\label{fig:SF2}
\end{figure}

\begin{figure}[h]
\centering
\subfloat
{\includegraphics[width=0.4\textwidth]{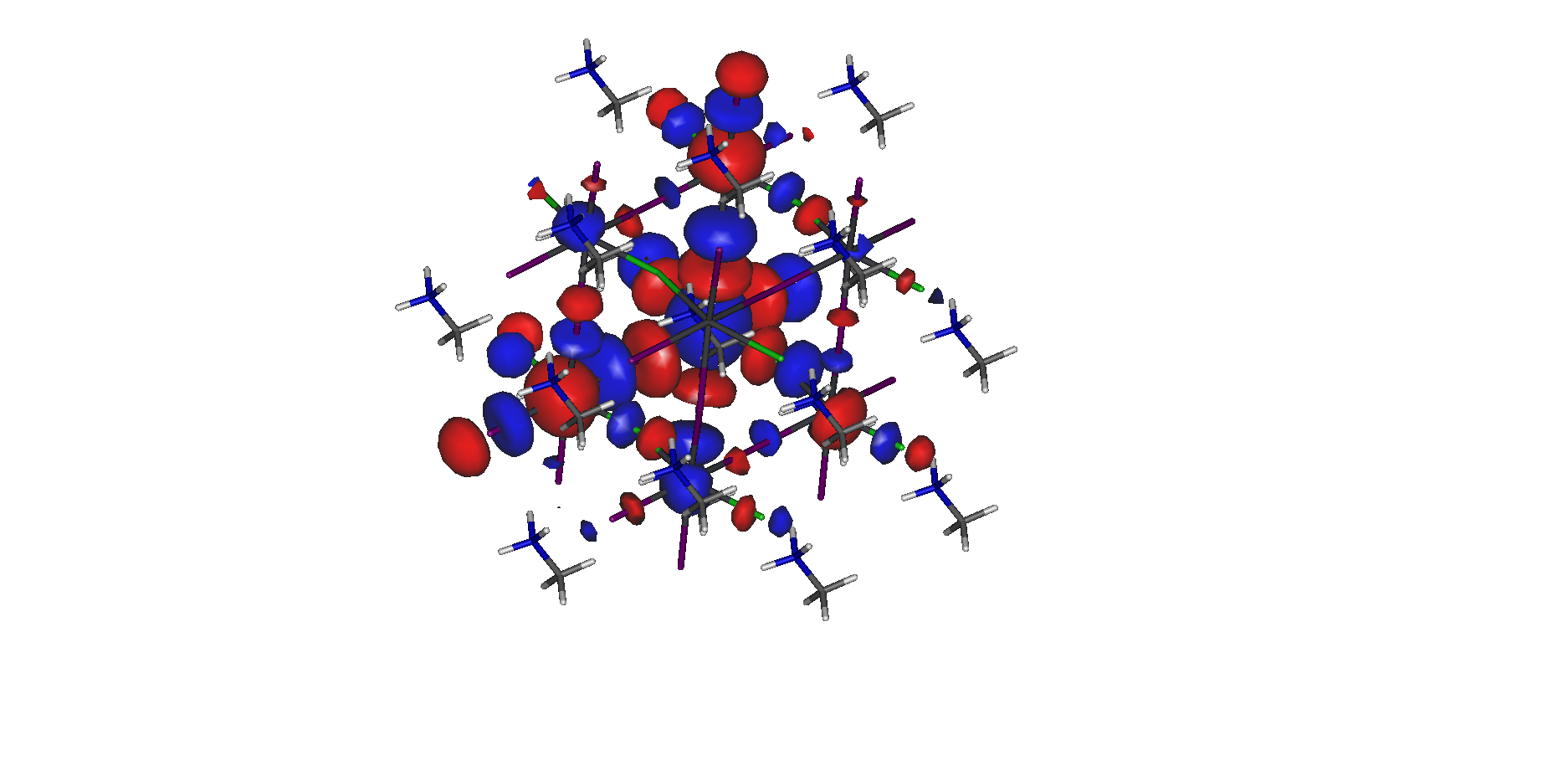}
\label{fig:SF3a}}
\qquad
\subfloat
{\includegraphics[width=0.4\textwidth]{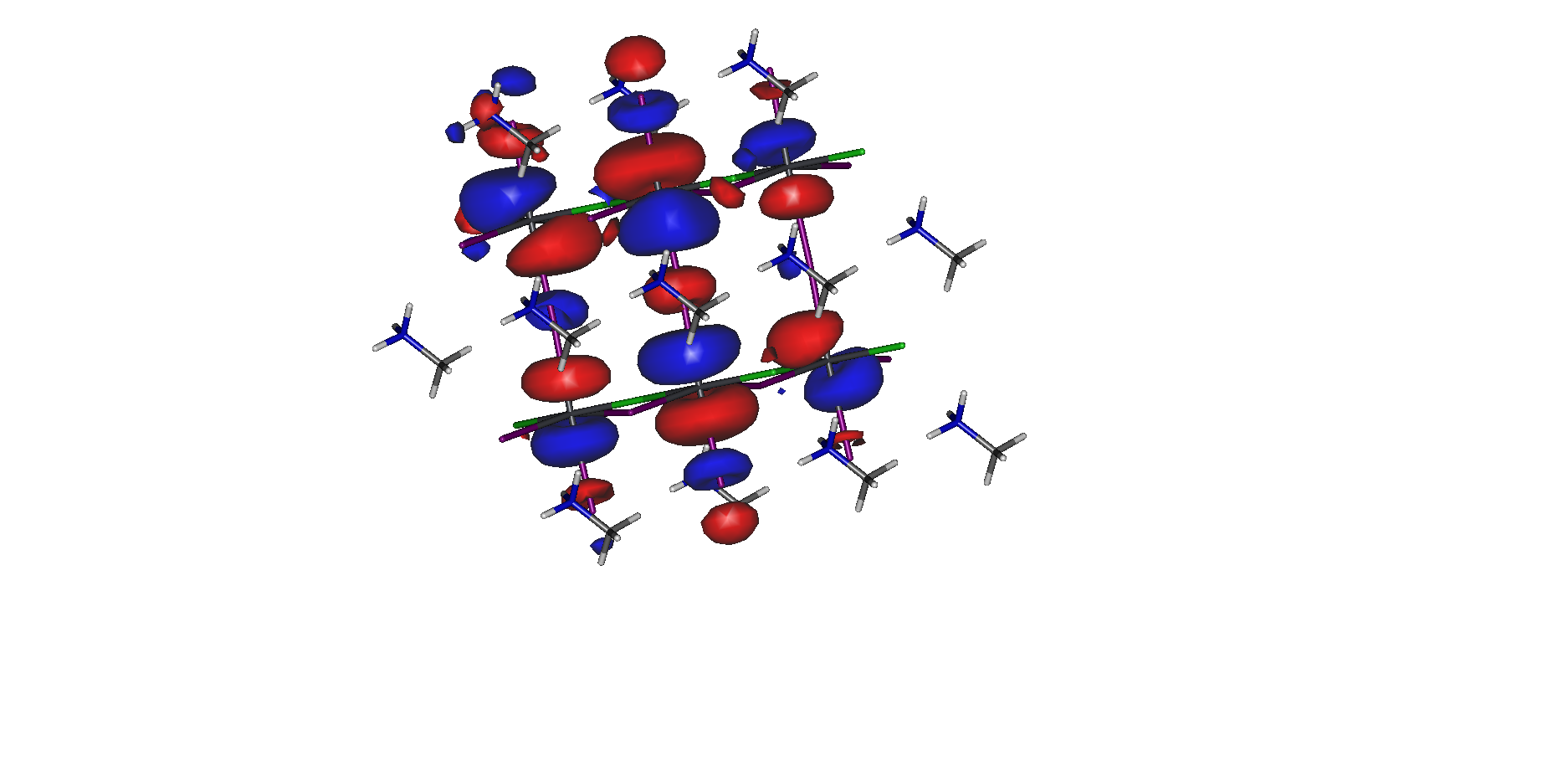}
\label{fig:SF3b}}
\caption{The natural transition orbitals (NTO) of the first excited singlet of energy 2.8 eV obtained from a finite cluster of 196 atoms using TDA/TDDFT; Left: The hole NTO of the first excited singlet has a Iodine p- and Lead s-character; Right: The electron NTO of the first excited singlet has a pronounced Lead p-character.}
\label{fig:SF3}
\end{figure}

\begin{figure}[h]
\centering
\subfloat
{\includegraphics[width=0.4\textwidth]{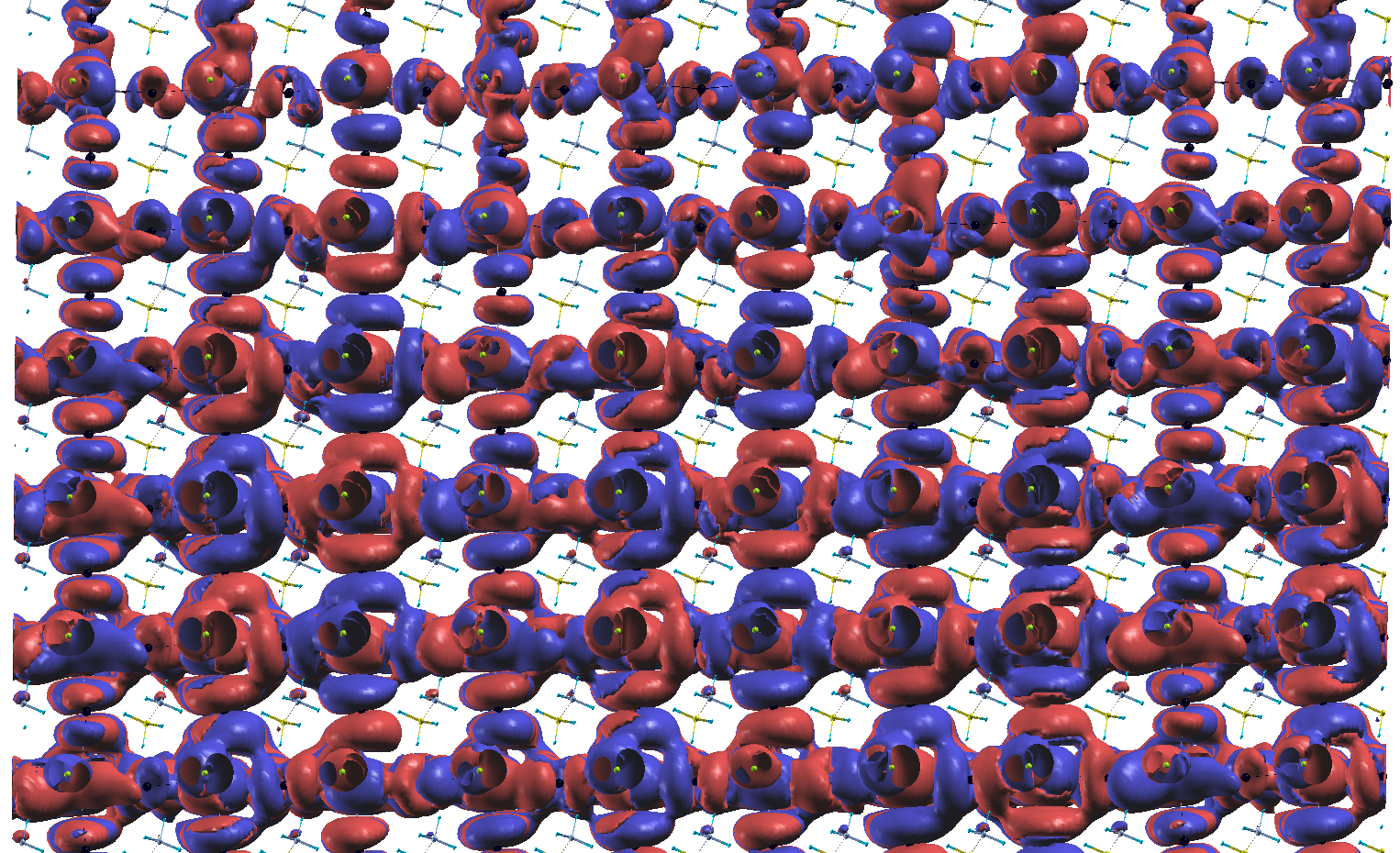}
\label{fig:SF4a}}
\qquad
\subfloat
{\includegraphics[width=0.4\textwidth]{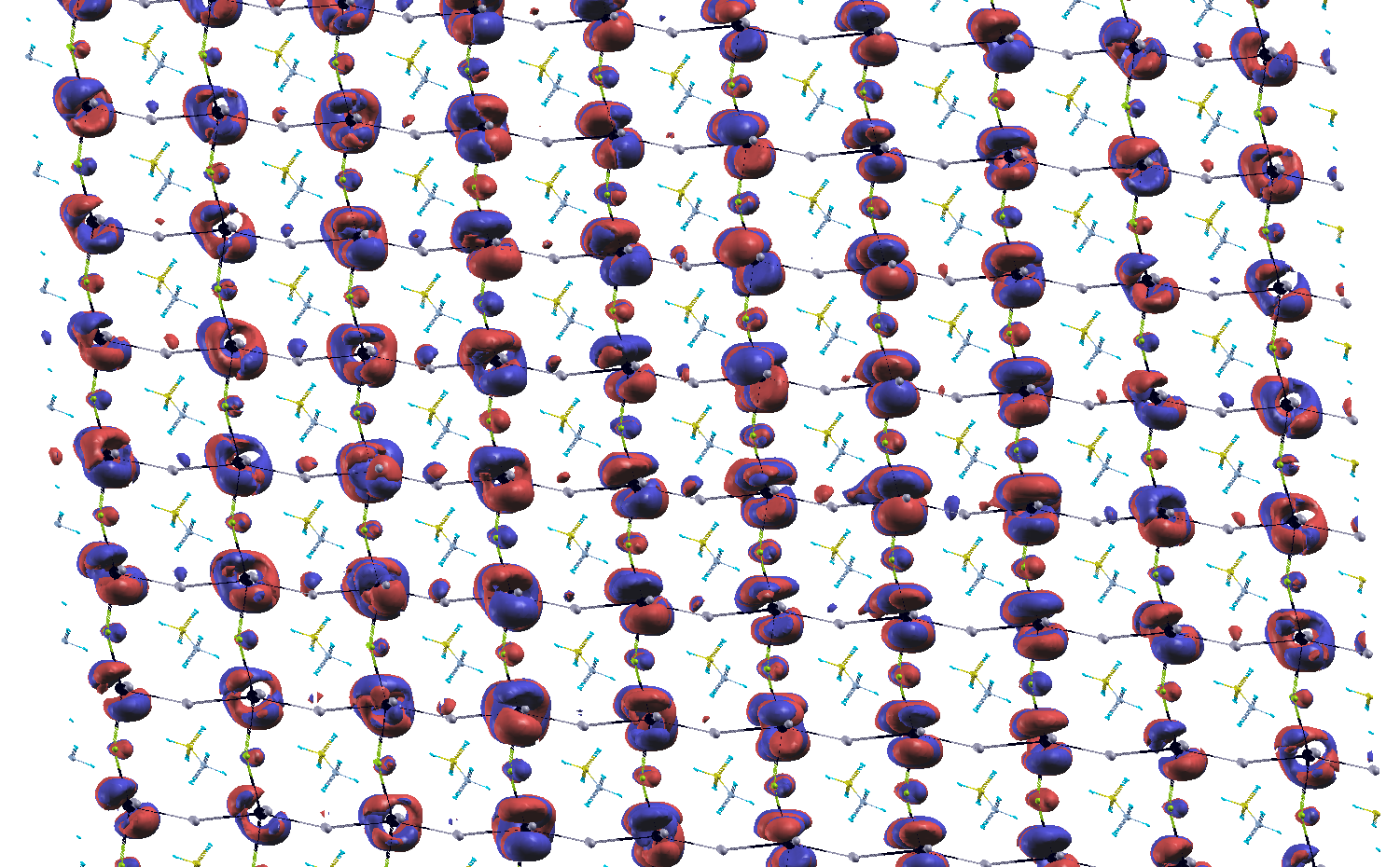}
\label{fig:SF4b}}
\caption{The hole and electron densities of the seventy-fifth singlet exciton, labelled $|2\rangle$ in the text, of energy 3.37eV obtained from the BSE; Left: The hole density has a Iodine p- and Lead s-character; Right: The electron density has a Lead p-character. The orbital characters are similar to those of exciton $|1\rangle$, the lowest exciton, but also has mixtures of other particle-hole transition states.}
\label{fig:SF4}
\end{figure}